\documentstyle[11pt,newpasp,twoside,epsf]{article}
\def\edcomment#1{\iffalse\marginpar{\raggedright\sl#1\/}\else\relax\fi}
\reversemarginpar

\begin{document}

\title{Photometric study of radio galaxies in RATAN--600 ``Cold'' survey}

\author{
    O.V.Verkhodanov,
    Yu.N.Parijskij,
    N.S.Soboleva,
    A.I.Kopylov,
    A.V.Temirova,
    O.P.Zhelenkova
}
\affil{Special Astrophysical Observatory, Nizhnij Arkhyz, Russia}

\author{W.M. Goss}
\affil{National Radio Astronomical Observatory, Socorro, NM, USA}

\begin{abstract}
   About 100 steep spectrum radio sources from the RATAN--600
   RC catalog were mapped by the VLA and identified with optical
   objects down to 24$^m$--25$^m$ in the R band using the 6\,m
   telescope. An updated list of calibrators with known redshifts
   of the same class RGs was compiled to evaluate the accuracy of
   photometric redshifts estimates.
   BVRI photometry for 60 RC objects was performed with
   the 6\,m telescope, and by standard model fitting we have
   estimated colour redshifts and ages of stellar population of
   host gE galaxies.
   The mean redshift of FRII RGs from the RC list happened to be
   $\approx$1.
   Several objects were found in which active star formation began in the
   first billion years after the Big Bang.
\end{abstract}

Catalogue of 1145 radio sources has been obtained in the RATAN--600
``Cold'' (RC) survey carried out in 1980--1988
with the radio telescope RATAN--600 in a strip of sky, $24^h$ long and
$0^\circ.6$ wide ($\delta \approx +5^\circ$), at 6 frequencies
from 1 to 22 GHz with a sensitivity of about 3 mJy at 3.9 GHz.

To select candidates to distant radio galaxies (RGs)
in the ``Big Trio'' project (RATAN$-$VLA$-$SAO~6\,m)
we used the most distant population of radio
sources in the 10--50 mJy range, where the $log(N/N_0)-log(S)$
curve has a maximum slope, and did
selection of about 100 steep spectrum (SS) radio sources ($\alpha\ge 0.9$).

At the following stages we did selection of the FRII type RGs
with the VLA maps
and optical identification and photometry with the 6\,m telescope.
65 objects look like FRII and about 20 of them belong
presumably to the most distant generation of RGs.
19 objects were classified as quasars by their stellar appearance
on CCD images.
Apart from the normal FRII RG class, some objects
with SS happened to be of quite different nature.
16 were not resolved even with VLA in ``A'' configuration at 3.7 cm,
being less than 0\arcsec.2 --1\arcsec.0 (CSS class).
16 ``subgalactic'' doubles, with sizes less
than 4\arcsec, are of separate interest.
A few of them are very complex and can not be considered as young
progenitors of FRII objects.

The technique of multicolour photometry has became in the past few
years as the main method in selecting candidates to distant galaxies,
and the only approach at very high redshifts. Determination of the age
of high redshift stellar systems may be the only way of estimation of
first galaxies formation redshift if star formation begins at redshifts
larger than $z$ of secondary ionization.
Direct observation of protogalaxies predicted by some recent computer
simulations is not possible.

The method of multicolour photometry was checked on a sample of 45
radio galaxies published in literature and carefully selected as steep
spectrum high $z$ objects of the FRII type (Verkhodanov et al., 1999).
It was shown that $BVRI$ colours are sufficient for accurate estimation
of $z$ and age in the redshift range of 0.5--3.5.

We have implemented this approach
for 60 FRII and CSS radio galaxies of the RC catalogue that
had been observed in $B,V,R_c,I_c$ bands
in 1994--1998 with CCD camera on the 6\,m telescope.
(Our sample of SS radio galaxies is now the largest one
with the four-band optical photometry.)
The data were used to estimate colour photometric redshifts
and ages of host galaxies by comparison with the PEGASE
(Fioc and Rocca-Volmerange, 1997)
and GISSEL'98 (Bruzual and Charlot, 1993) models of evolution
of spectral energy distribution (SED).
For few typical cases the reality of the colour
$z$ determinations was confirmed spectroscopically (Dodonov et al., 1999).

\begin{itemize}
\item
      It was shown photometric redshifts give tolerable (10--20\% err)
      agreement with spectral ones for powerful RGs.
\item
      A redshift--magnitude diagram shows much larger dispersion
      at $m_R$$\ge$22$^m$.
\item
      Distribution by $z$ of our objects has a maximum
      at $z\sim 1$, i.e. near the maximum Universe activity stage,
      and $\sim$10 objects at $z$$>$2.5.
\item
      A galaxy age is model-dependent and detected uncertainly
      (in limits of 0.5--10\,Gyr).
      One can point a low limit of galaxy age and, since, $z$ of
      its formation.
      This age is always larger than standard estimation of
      radio source lifetime.
\end{itemize}

The mean multicolour $(BVRI)$ age of the stellar population of
the RC radio source host
galaxies is of order of 1 Gyr, and at least in some objects active star
formation had begun in the first Gyr after the Big Bang. Such distant
objects must be of high density contrast and
modern cosmology has to explain this very early appearance of dense
and massive (Tera--solar masses) protogalaxies, with quickly formed massive
black holes inside them to produce FRII structures.
>From our 100 square degree area sample we can estimate
that more than 10000 very early objects, born before the QSO epoch, are
available on the sky and accessible for present-day optical and radio
facilities. They can help us to penetrate into the ``Dark Age'' of the
Universe, between the recombination epoch and the epoch of
appearance of QSOs.

  Future activity connected with the ``Big Trio''
project will be concentrated  on the direct spectroscopy of the most
probable candidates to the first galaxy generation selected from
our RC list.

\end{document}